
\input harvmac

\input epsf
\ifx\epsfbox\UnDeFiNeD\message{(NO epsf.tex, FIGURES WILL BE IGNORED)}
\def\figin#1{\vskip2in}
\else\message{(FIGURES WILL BE INCLUDED)}\def\figin#1{#1}\fi
\def\ifig#1#2#3{\xdef#1{fig.~\the\figno}
\goodbreak\midinsert\figin{\centerline{#3}}%
\smallskip\centerline{\vbox{\baselineskip12pt
\advance\hsize by -1truein\noindent\footnotefont{\bf Fig.~\the\figno:} #2}}
\bigskip\endinsert\global\advance\figno by1}

\Title{\vbox{\baselineskip12pt\hbox{hep-th/9211030}\hbox{RU-92-40}}}
{\vbox{\centerline{Black Hole Remnants and the Information Puzzle}
}}
\def\ajou#1&#2(#3){\ \sl#1\bf#2\rm(19#3)}
\lref\gm{G. Gibbons and K. Maeda, \ajou Nucl.
Phys. & B298 (88) 741.}
\lref\ghs{D. Garfinkle, G. Horowitz, and A.
Strominger, \ajou Phys. Rev. &D43 (91) 3140; Erratum: \ajou Phys. Rev.
&D45 (92) 3888.}
\lref\BDDO{T. Banks,
A. Dabholkar, M.R. Douglas, and M. O'Loughlin, ``Are
horned particles the climax of Hawking evaporation?'' \ajou Phys. Rev.
&D45 (92) 3607.}
\lref \RST{J.G. Russo, L. Susskind, and L. Thorlacius, ``
Black hole evaporation in 1+1 dimensions'' \ajou Phys. Lett. &B292 (92) 13.}
\lref\dxbh{S. B. Giddings
and A. Strominger, ``Dynamics of extremal black holes'',\ajou Phys. Rev.
&D46 (92) 627.}

\centerline{T. BANKS and M. O'LOUGHLIN}
\baselineskip18pt
\centerline{\sl Dept. of Physics and Astronomy}
\centerline{\sl Rutgers University}
\centerline{\sl Piscataway, NJ 08855-0849}
\smallskip
\centerline{\sl and}
\smallskip
\centerline{ANDREW STROMINGER}
\baselineskip18pt
\centerline{\sl Dept. of Physics and Astronomy}
\centerline{\sl Rutgers University}
\centerline{\sl Piscataway, NJ 08855-0849}
\centerline{\sl and}
\centerline{\sl Department of Physics}
\centerline{\sl University of California}
\centerline{\sl Santa Barbara, CA 93106-9530}
\smallskip
\noindent
Magnetically charged dilatonic black holes have a perturbatively
infinite ground state degeneracy associated with an infinite volume
throat region of the geometry.  A simple argument based on causality is
given that these states do not have a description as ordinary massive
particles in a low-energy effective field theory.  Pair production of
magnetic black holes in a weak magnetic field is estimated in a
weakly-coupled semiclassical expansion about an instanton and found to
be finite, despite the infinite degeneracy of states.  This suggests
that these states may store the information apparently lost in black
hole scattering processes.

\vskip1in
\eject

A semiclassical analysis \ref\hawkfirst{S. W. Hawking,\ajou Commun.
Math. Phys. &43 (75) 199 \semi \ajou Phys. Rev. &D14 (76) 2460.} implies
that  the radiation
emitted by an evaporating black hole is not in a pure state:  it is
correlated with radiation which falls into the black hole.  Causality
apparently prevents one from retrieving this information until the black
hole is Planck-sized and the semiclassical approximation has broken
down \ref\inf{See for example, R.M. Wald, ``Black holes, singularities,
and predictability," in Quantum Theory of Gravity, S.M. Christensen ed.
(Adam Hilger, Bristol U.K. 1984), or J. Preskill, ``Do black holes
destroy information?" Caltech preprint CALT-68-1805, hep-th/9209058.}.
The amount of
information that can possibly be retrieved by an
external observer (in any reasonable time period) is then
limited by energy conservation.  There are limits to how much
information can be carried out in a finite time by radiation with  a
finite total energy \ref\cash{Y. Aharonov, A. Casher and S. Nussinov,
\ajou Phys. Lett. &191B (87) 51.}.  One possibility is that the information
is simply lost \hawkfirst.  Another is that the black hole stops
evaporating when it reaches the Planck mass, and the information is
stored in a stable (or very long lived) remnant (see for example
\refs{\hawkfirst,\cash}).
The remnant must have an essentially infinite number of quantum states
in order to accomodate the information.  This second possibility has
been severely criticized on the grounds that such an infinite degeneracy
of remnants would show up as an infinite contribution to
real or virtual processes.

A simpler problem, for which an analogous information puzzle occurs, is
the scattering of low-energy particles by extremal black holes.  This
occurs via classical absorption followed by Hawking reemission, and
information appears to be lost in the process unless the extremal
black hole has an infinite ground state degeneracy.

In fact, it was noted in \ref\cghs{C.G. Callan, S.B. Giddings,
J.A. Harvey, and A.
Strominger, ``Evanescent black holes," \ajou Phys.
Rev. &D45 (92) R1005.} that recently studied \refs{\gm,\ghs}
extremal magnetic black holes in
dilaton gravity appear to have just such an infinite degeneracy
associated with an infinite volume throat region of the
geometry\foot{This infinite degeneracy might be removed by
non-perturbative effects \refs{\BDDO,\RST}, as will be briefly discussed.  In
this paper we assume the degeneracy exists and investigate its
consequences.}.
In this paper we will compute the production rate of these states (which will
be referred to as ``cornucopions''\BDDO) in a weak magnetic field.
For sufficiently weak coupling  the rate can be reliably computed in a
semiclassical expansion about an instanton describing the creation of a
geometry of finite size.  We find that the rate is
finite, despite the infinite degeneracy.  We are able to
relate the finite volume of the states produced by tunneling to a
horizon which is produced a finite distance down the cornucopion in
the process of moving its throat around in the external spacetime.
The external field determines both the distance to the Minkowski
horizon, and the size of the instanton. This
reconciles the euclidean and minkowskian descriptions of the infinite set
of cornucopion states.  They cannot be treated as elementary particles
in either regime, although the classical positions of their throats
follow particlelike trajectories.

It was previously argued in \ref\corn{T. Banks and M. O'Loughlin,
``Classical and quantum production of cornucopions at energies below
$10^{18}$ GeV'', Rutgers preprint RU-92-14 (1992).} that the production rate of
magnetically charged black holes in dilaton gravity is finite.  There it was
suggested that the tunneling process which is responsible for
cornucopion production produces an initial state of finite volume which
then expands classically to become the essentially infinite cornucopion
volume.  As in
the inflationary universe, the number of states of the final geometry
that can be produced in this way is limited by the size of the initial
geometry.   The present paper confirms the final conclusions of \corn ,
but the physical picture developed here is quite different than that envisaged
there.

While our results do not directly bear on the information puzzle
for neutral black holes, they certainly suggest that the standard
arguments against remnants may be wrong in the neutral case as well.
In the conclusion we
will argue that the cornucopion hypothesis is really a hybrid between
the hypotheses of remnants and information loss, which avoids the
difficulties of both.

Four-dimensional dilaton-Maxwell gravity is described by the action
\eqn\ctn{S = {1 \over 16\pi}\int d^4 x\sqrt{- g}e^{-2\phi}(R + 4(\nabla
\phi)^2 - {1\over 2}F^2).}

\noindent
This theory has a three-parameter family of magnetic black hole
solutions labelled by the mass $M$, magnetic charge $Q$ and asymptotic value
$\phi_0$ of the dilaton \refs{\gm, \ghs}.  For $M^2$ greater than the
critical value ${1 \over4} Q^{2}$, the causal structure of these black holes is
similar to that of the Schwarzschild geometry,
with a singularity behind an event horizon.  For $
M^2 < {1 \over 4}Q^2$, the solutions resemble $M^2 <
Q^2$ Reissner-Nordstrom:  there is a naked singularity at the origin.

\ifig\fone{The spatial geometry of an extremal dilatonic magnetic black hole.
The cross sections of the throat are two spheres.}
{\epsfysize=3.0in\epsfbox{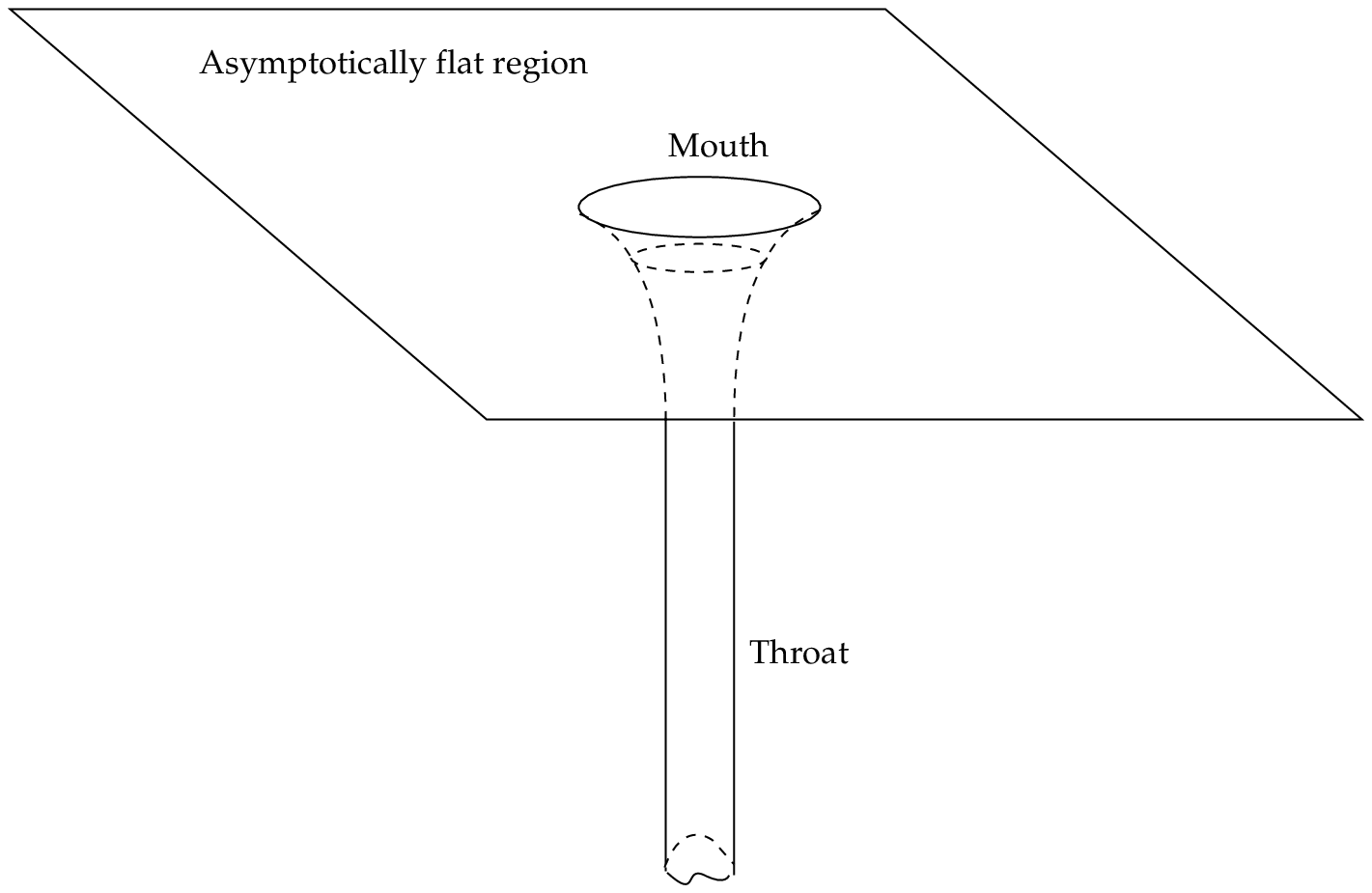}}

The nature of the extremal $M^2 = {1\over4}Q^{2}$ solutions
is, however, dramatically affected by the presence of the dilaton \ghs .  The
extremal solution is given by
\eqn\xbh{\eqalign{ds^2 &= -dt^2 + (1 + {Q \over y})^2
(dy^2 + y^2 d^2\Omega),\cr e^{2(\phi - \phi_0)} &= 1 + {Q \over y},\cr
F &= Q \epsilon,}}
where $\epsilon$ is the volume form on the two sphere of constant radius,
normalized so that its  integral is equal to $4\pi$.  As
illustrated in Figure 1, the spatial geometry is asymptotically flat for
large $y$, but as $y$ approaches zero, there is a semi-infinite ``throat''
whose cross sections are two spheres of constant radius $Q$.  The region
$y\sim Q$, where the throat begins, will be referred to as the mouth.  The
extremal geometry is geodesically complete, and there are no horizons or
singularities.  However there are extra null infinities, since light
rays can travel forever down
the throat.  The Penrose diagram is depicted in Figure 2.  The dilaton field
grows linearly in this region, which means that gravity becomes strongly
coupled far down the throat.

\ifig\ftwo{Penrose diagram of an extremal magnetic dilatonic black hole.}
{\epsfysize=3.0in\epsfbox{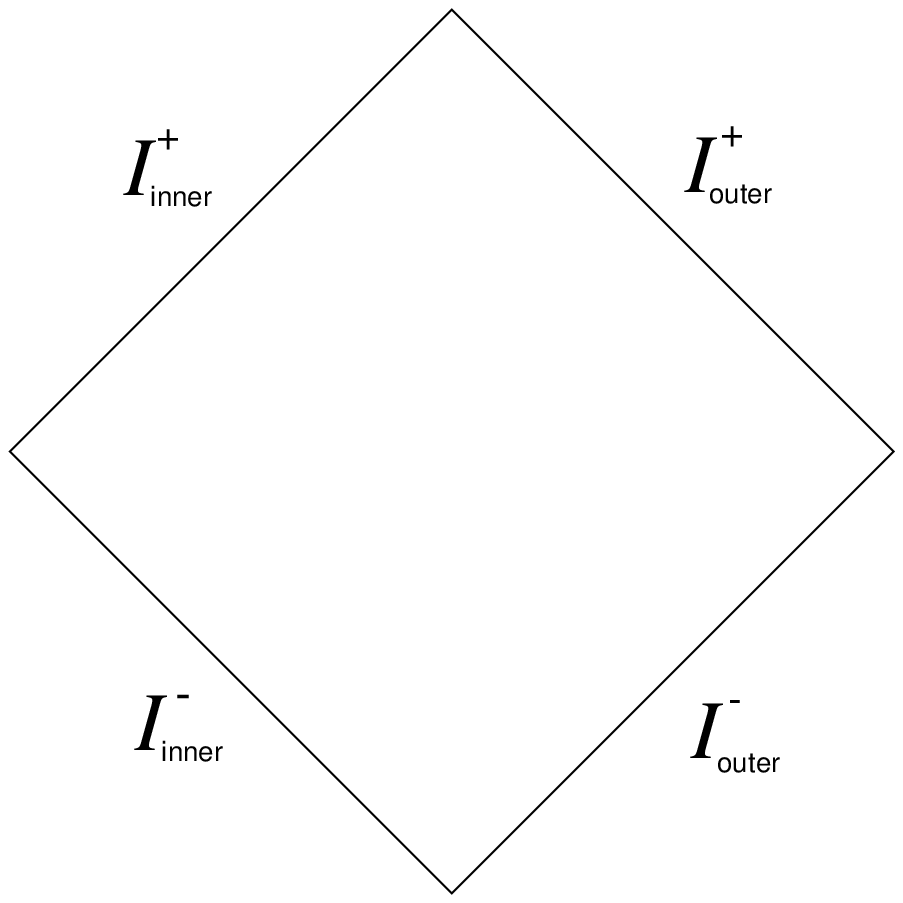}}

Recent attention has focused on the problem of low-energy $S$-wave
scattering of massless scalars or charged fermions by this extremal
black hole \cghs .
As soon as any energy is thrown into the throat, a horizon develops and
the black hole becomes non-extremal.  It then presumably returns to its
extremal ground state via Hawking emission.  At energies much less than ${1
\over Q}$, this process can be analyzed with the two-dimensional
effective field theory which governs low-energy dynamics in the throat
region.  Regarding this region as a compactification from four to two
dimensions
with a two sphere threaded by magnetic flux, the two-dimensional effective
field theory is given by\foot{The overall factor of $Q^2$, arising
from the volume of the two spheres, was absorbed into a shift of $\phi$
in \refs{\BDDO, \dxbh}.  $Q$-independent factors have been absorbed into a
shift  of $\phi$ in (3).}
\eqn\ttn{S_2 = {Q^2 \over 2\pi}\int d^2 x\sqrt{- g}e^{-2\phi}(R+4(\nabla
\phi)^2 + {1 \over Q^2} - {1\over2}G^AG^A)}
together with the appropriate matter
action. $G^A$, $A=1,2,3$ are $SU(2)$ gauge field strengths arising from the
isometries of the two sphere.  Although
mentioned in \dxbh , these modes have been ignored in previous work because
rotational invariance prevents them from being excited in $S$-wave scattering.
They will however play a major role in the rotationally non-invariant processes
considered later in this paper.  We are suppressing in  \ttn\ the relic of the
original gauge field {\it F} as it will not enter into later considerations.
The theory \ttn\ has a vacuum solution
\eqn\ldv{\eqalign{g_{ab}&=\eta_{ab},\cr
\phi &= -{x \over {2 Q}},}}
which corresponds to the extremal four-dimensional
magnetic black hole.  There are also two-dimensional black hole
solutions \ref\wittTwod{E. Witten, ``On string theory and black holes,''
\ajou Phys. Rev& D44 (91) 314.} which correspond to non-extremal
four-dimensional magnetic
black holes.
Stationary solutions with non-zero $G$
correspond to the charged rotating black hole solutions of
\ref\rbh{A.Sen, ''Rotating charged black hole solution in heterotic
string theory'', \ajou Phys. Rev. Lett. &69 (92) 1006, J. Horne and G.
Horowitz, ``Rotating dilaton black holes'', \ajou Phys. Rev. &D46 (92)
1340.}.
Four-dimensional particle-hole scattering corresponds to two
dimensional black hole formation/evaporation.

In \cghs\ semiclassical
equations describing this process for a large number $N$ of matter fields
were derived.  It was conjectured that it is
described by a unitary $S$-matrix, with the infinite amount of information
about the incoming scattering state being stored in an infinite number
of zero-energy bound states of the extremal black hole (i.e.
cornucopions).  The possibility
of an infinite degeneracy of states arises because of the infinite volume of
the throat region.  This conjecture was shown to be false for large $N$  in
 \BDDO\ and \RST .  Rather a singularity is formed and the approximation
breaks down.
However in \ref\Stro{A. Strominger, ``Fadeev-Popov ghosts and 1+1
dimensional black hole evaporation,'' {\it Phys. Rev. D}, to appear,
hep-th /9205028.} evidence was presented that, after accounting for the
effects of  Fadeev-Popov ghosts, the conjecture of \cghs\  might be realized
when $N < 24$.

We shall henceforth assume that, at least in some model,
particle-hole scattering is indeed unitary and the information is accounted
for by the infinite set of cornucopions.  Whether
or not this is the case is an important question that is not the
subject of this paper.  Rather, we wish to question the following
argument that any remnant scenario of this type is in fact experimentally
ruled out.

In a low-energy effective field theory, the argument goes, these degenerate
cornucopions are described by an effective action
\eqn\cnt{S_{eff}= \sum_{i=1}^\infty \int d^4x \sqrt{-g}e^{-2\phi}
({\cal D}_\mu \Psi_i {\cal D}^\mu \Psi_i +
{Q^2\over4}\Psi_i^2),}
where $\Psi_i$ creates the $i$th cornucopion; and $\cal{D}$ is the
(somwhat complicated) covariant derivative for a magnetically charged particle.
The low energy field theory contains instantons, depicted in Figure 3,
describing cornucopion pair production in a weak magnetic field.  It is
usually asserted that as long as
the magnetic field is small compared to the cornucopion mass $M=Q/2$, and the
Schwinger length $M/QB$ much larger than the diameter of the
throat, these
instantons should give an accurate estimate of the production rate.
According to Schwinger, the production rate (with our conventions) is
proportional to
\eqn\prt{exp[{-{\pi Qe^{-2 \phi_0}} \over 4B}]}
for each species of cornucopion. $e^{2\phi_0}$ plays the role of
$\hbar$ in the
exponent. For $Q$ of order one (in Planck units) and
ordinary magnetic fields, \prt\ is a fantastically small number.
However we must sum over all species
of cornucopions,
which yields an infinite total production rate!  Since cornucopion
production  has not been
observed, one concludes that there cannot possibily be enough states of
an extremal black hole to account for all the potential information loss
in scattering experiments.

\ifig\fthree{The instanton for pair production of charged particles in a
magnetic field is a circular Euclidean
orbit.}{\epsfysize=2.0in\epsfbox{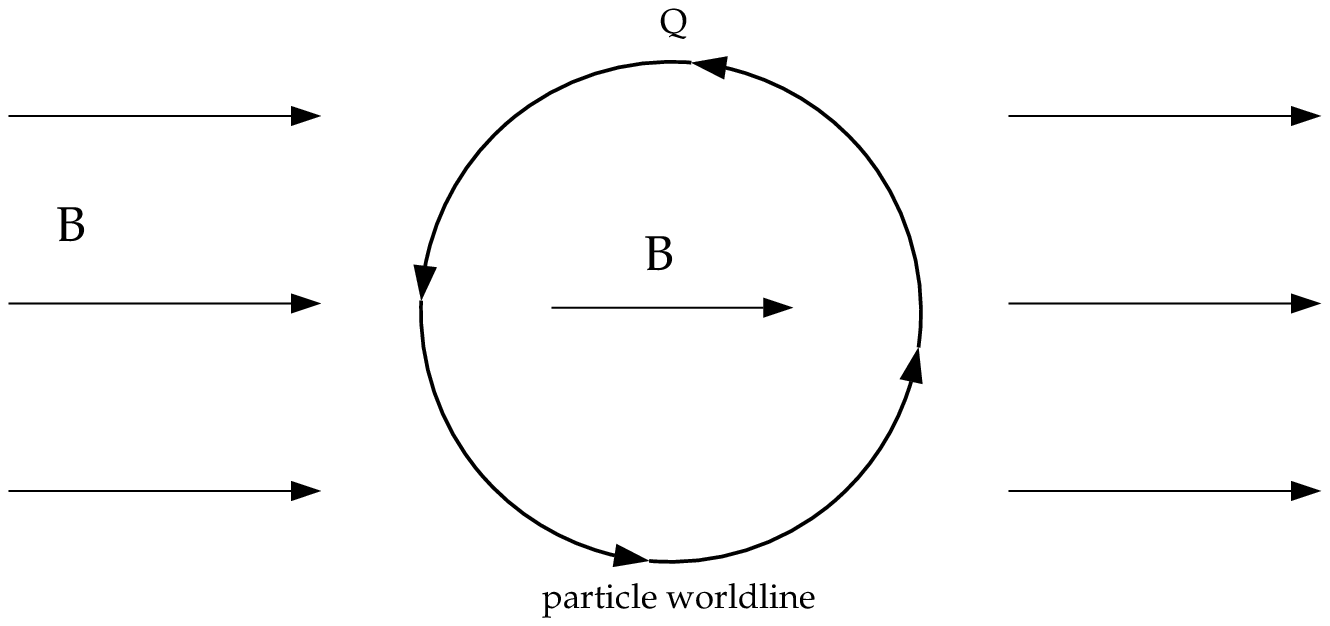}}

This argument is false.  In fact cornucopions can not be simply described
by an effective action of the type \cnt.  To sharpen this discussion,
it is useful to consider a theory in which the throat has a finite, but
very long length, say a million light years \foot{Such a theory can be
constructed by replacing the gauge field in \ctn\  with a GUT theory broken
to $U(1)$.  For a low GUT scale, the theory has `tHooft-Polyakov monopoles
with small gravitational corrections.  For a GUT scale greater than the
Planck scale, the magnetically charged solutions are identical to \xbh. By
tuning the GUT scale to the approach the Planck scale from below, one
obtains an arbitrarily long throat terminated at a monopole.}.  There
will then be a large but
finite, say ${\cal{N}} \sim e^{10^{6} \times 10^{7} \times 10^{42}}$,
number of distinguishable cornucopions.

Now let us suppose we are given a box containing $10 {\cal{N}}$ cornucopions
and
asked to sort them according to species into ${\cal{N}}$ boxes.  Assuming they
are described by an effective action of the type \cnt, and given a large
array of very well-equipped experimentalists, this is a straightforward task.
One simply performs interference experiments to determine which
cornucopions are identical
and then drops them in the boxes.

This experiment is doomed to failure if it is performed in less than a
million years.  The reason is simple.  If we have fully determined the
species of the extremal black hole, we have obtained information about the
quantum state in a region of spacetime a million light years away.
Causality clearly must prevent us from obtaining this information.
We emphasize that what we have described is a true violation of
causality, and not an Einstein-Podolsky-Rosen paradox.  The operation of
moving the cornucopions around within a room in a matter of minutes or
hours can be described by the action of local operators in a region of
spacetime that is causally separated from the bottom of the throat during
the same time interval.  If our experiment allowed us to
 determine whether two states of the
cornucopion localized at the bottom of the throat were the same or not,
these operators would not commute with a local operator which changed
the state near the bottom of the throat during the same time interval.
We conclude that \cnt\  can not provide a full description of the low-energy
dynamics of cornucopions; and we accordingly have no right to conclude that the
cornucopion production rate diverges.

What goes wrong when we actually try to do the experiment?  In order to
interfere two cornucopions, we must move them.  Some energy will
necessarily go down the throat, and a horizon will form.  This is the
second law of black hole mechanics:  interactions with a black hole
increase the area of the horizon.  The more gently the cornucopions are
moved, the farther down the throat the horizon is formed.  If they are
 moved extremely gingerly (on a time scale exceeding a
million years) a horizon may not form at all - but this is not possible
for experiments which take less than a million years.  Once a horizon
forms all bets are off.  Presumably it eventually recedes via Hawking
emission, but the state of the cornucopions is greatly perturbed by the
attempt to measure it. Even if they were initially in identical states
the probability that they remain so after the scattering is completed is
very small.  The formation of horizons in the internal geometry
prevents us from deciding whether the cornucopions are bosons, fermions
or distinguishable particles as long as the length and time scales
involved in the scattering process are short compared to the cornucopion
length.  If this length is finite, a naive effective field theory
analysis would be valid only at much longer length scales.

These observations can be quantified in the context of
the pair production of magnetic black holes in a weak magnetic field.
For the Einstein-Maxwell theory without dilatons, this production rate
was  computed using
instanton methods in
\ref\gast{D. Garfinkle, A. Strominger, ``Semiclassical Wheeler wormhole
production'' \ajou Phys. Lett. &B256 (91) 246.}.  The instanton -in
analogy to that discovered by Schwinger for point particles -
is the analytic continuation of a real time
configuration describing two constantly accelerating, oppositely charged
dilatonic magnetic black holes in a weak magnetic field.  As depicted in
Figure 4, along the hyperbolic wordlines of the two black hole mouths  are
attached two ``fins'' corresponding to the throats of the black holes.
We depict these fins as having finite length because the application of
an external
force on the extremal black hole produces a horizon a finite distance
down the throat\foot{ The
formation of a horizon may also be understood as a requirement for the
black hole to thermally equillibrate with the acceleration radiation.}.
The geometry of the uniformly accelerating cornucopions has a
timelike Killing field which generates motion along the hyperbolae and is
null (zero norm) at the horizons.

\ifig\ffour{Uniform acceleration of two oppositely charged black holes
in a magnetic field B. The throats are terminated at the horizon.}
{\epsfysize=2.1in\epsfbox{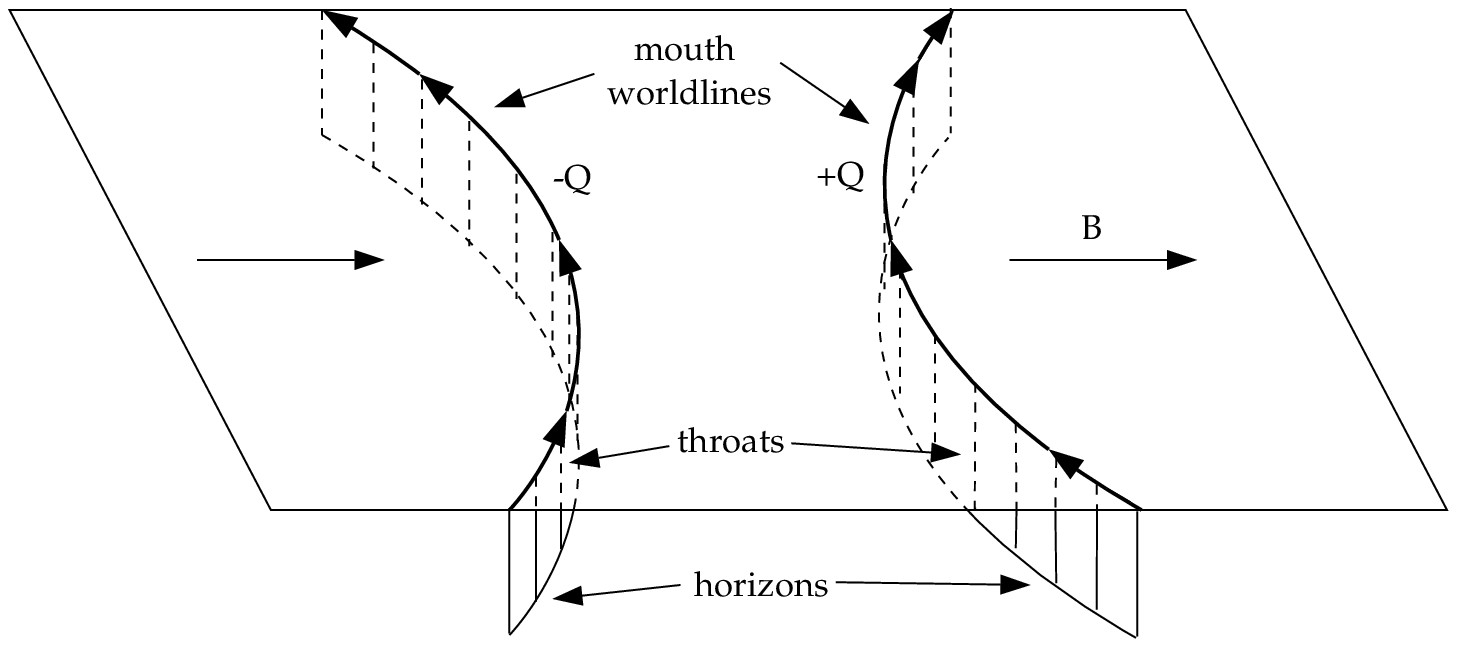}}

The analytically continued euclidean geometry is depicted in Fig. 5.
The hyperbolae become circles, and the Killing field generates a
rotational symmetry.  This symmetry has a fixed point on the horizons,
which collapse to a single point, closing up the end of the fins. Thus,
as in previous examples of gravitation instantons, the euclidean section
is a continuation of only the region of the Minkowski geometry outside
the horizon.   The
result is that the euclidean continuation of the
throat region has the shape of a cup\foot{In our effective field theory
the two-spherical cross section of the throat is replaced by a point,
and the world surface of this infinitely thin throat is the surface of
the cup.  We also note that the solutions of the field equations have
negative curvature, so that Fig. 5, and the term {\it cup} may be
somewhat misleading.  We have not been able to find graphics software
capable of drawing a faithful image of the geometry on a sheet of
paper.} which is joined
onto the asymptotically flat region along its rim, the worldline of the black
hole mouth.

\ifig\ffive{Instanton for magnetic pair creation of charged black holes.
A finite volume cup-shaped region is attached to the asymptotic region
along the circular mouth worldline.}
{\epsfysize=2.5in\epsfbox{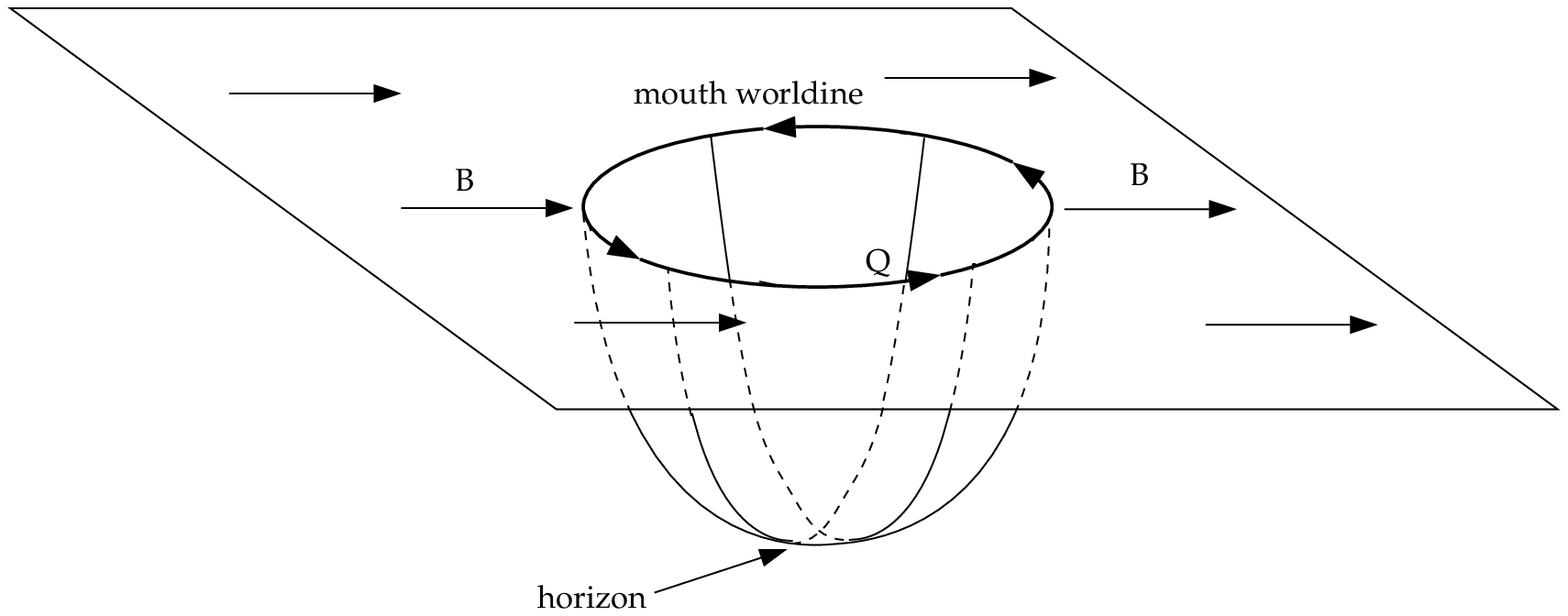}}

The exact instanton is known for Einstein-Maxwell theory \gast.  It is
certainly
difficult, but perhaps possible, to construct the analogous solution in
the dilaton-modified theory.  However at scales below ${1 \over Q}$
one may use the novel effective field theory described in
\refs{\cghs, \BDDO, \dxbh} to describe the instanton.  This effective field
theory has mixed dimensionality.  There is a four
dimensional part in which the black
holes are described as charged point particles moving in weak magnetic,
gravitional and dilatonic fields.  We shall consider only magnetic
fields, so the relevant part of the four-dimensional action is (for a
single black hole)
\eqn\noname{S_4= {1 \over 32 \pi} e^{-2\phi_0} \int d^4xF^2 +
\half e^{-2\phi_0}Q \int d\tau \tilde{A}_\mu (X)\dot{X}^\mu + e^{-2\phi_0}
m \int d\tau
\sqrt{\dot{X}^\mu \dot{X}_\mu},}
where $X^\mu(\tau)$ is the world line of the black hole mouth and - because the
charge is magnetic - $d \tilde{A}$ is the dual of the field strength.

In the effective field theory,
the mouth worldline is attached to a two-dimensional fin of possibly
infinite depth.  The dynamics of the fin are governed by the
two-dimensional dilaton gravity action \ttn.  In addition there are
interaction terms involving products of two and four-dimensional
operators integrated along the mouth worldline.  These allow energy
and information
exchange between the four and two-dimensional regions.  In general one
expects every interaction allowed by symmetries to appear.  One such relevant
term is ( in a non-relativistic approximation, valid for small $B^A$)
\eqn\tfc{ {i \gamma \over \pi} e^{-2\phi_0} Q \int H^A(0,\tau)B^A(X(\tau)),}
where $B^A$ is the four-dimensional magnetic field, $H^A(0,\tau)$ is
the two-dimensional SU(2) one-form potential  at the boundary
of the two-dimensional region.  We have not calculated $\gamma$ but expect
that it is order one. \tfc\ is gauge invariant because $B^A$ carries
SU(2) charge, i.e. it transforms under rotation.  \tfc\ implies that an
external magnetic field
will excite one of the SU(2) fields, and  that if $H^A$ is
non-zero on the boundary of the two-dimensional region, the black hole
has a magnetic dipole moment.

We now describe the instanton for black hole pair production in the
language of this effective field
theory.  The four-dimensional portion consists of a weak magnetic field
in euclidean $R^4$ with the mouth worldline moving in a circular orbit of
radius R.  As noted above, the two-dimensional portion is a cup whose
rim is joined to
the $R^4$ along the mouth worldline.  The geometry of this cup is a
truncation of the extremal, charged euclidean black hole solution
discussed in \ref\cbhs{M. McGuigan, C. Nappi and S. Yost,``Charged
black-holes in two-dimensional string theory'', {\it Nucl. Phys.}{\bf
B375} 421 (1992); G.W. Gibbons and M.J. Perry, ``The physics of 2d stringy
spacetime'', preprint hep-th/9204090 (1992); C. Nappi and A.
Pasquinucci, ``Thermodynamics of two-dimensional
black-holes'',  gr-qc/9208002 (1992).}:
\eqn\none{\eqalign {ds^2 &= Q^2 \lbrack {\sinh^2 r \over {(\cosh \alpha
+\cosh r)^2}} d\tau^2+dr^2],\cr e^{-2\phi}&=e^{-2\phi_0} {(\cosh \alpha
+ \cosh r)\over {(\cosh \alpha + \cosh r_M)}},\cr G &= {i\epsilon_2
 \sinh \alpha
\over {Q(\cosh \alpha + \cosh r)}},}}
where
\eqn\last{\eqalign{&0 < \tau <2\pi(\cosh \alpha + 1), \cr
&0 < r < r_M,}}
($\epsilon_2$ is the volume form on two space), and $r_M$ obeys
\eqn\lastagain{R= {Q \sinh r_M (\cosh \alpha +1) \over {(\cosh \alpha +
\cosh r_M)}}.}
insuring that the boundary length is $2 \pi R$.  $G$ is an abelian field
strength in a subgroup of $SU(2)$ (determined by the external $B$ field) and
$\alpha$ parameterizes its magnitude.

The parameter $\alpha$ is determined by  matching conditions as follows.
The boundary conditions on the two-dimensional field theory (see \dxbh\
for a detailed discussion) require that $G$ vanish if the operator
insertion \tfc\ is regarded as just inside the boundary.  Thus the
conserved black hole charge
\eqn\insertB{q= - * e^{-2\phi}G \mid_{r_M},} ($*$ is the Hodge dual) must
equal minus the charge $B \over Q$ near the boundary:
\eqn\qb{q=-  {{i e^{-2 \phi_0} \sinh \alpha }  \over {Q (\cosh \alpha +  \cosh
r_M)}}= - {i\gamma B e^{-2\phi_0} \over Q}.}
We are interested in large $R$, (small $B$), which implies large $r_M$.
One then has
\eqn\sha{\sinh \alpha \approx \gamma B \cosh r_M,}
or using (11)
\eqn\ftt{\eqalign{\cosh \alpha &\approx {R \over Q}, \cr \sinh r_M
&\approx {R \over {\gamma Q B}}.}}
Note that the correct euclidean $SU(2)$ field strength is imaginary,
as appropriate for instantons which describe tunneling to configurations
with non-zero electric fields \ref\lee{K. Lee, ``Wormholes and
Goldstone bosons"
 \ajou Phys. Rev. Lett. &61 (88) 263}.
The external magnetic field carries real Minkowski angular momentum, so
its effect on the two-dimensional Euclidean action is to insert a Wilson
loop on the boundary of the cup.  The semiclassical field due to a
Wilson loop is always imaginary.

It remains to extremize the entire four plus two-dimensional action with
respect to $R$.  The two-dimensional action contributes only a surface
term along the rim of the cup, which can be absorbed (for small $B$)
by a renormalization of the mass in \noname\ . It is easily seen that
small $B$  implies large $R$  and the
problem reduces to the one analyzed by Schwinger.  One finds that
\eqn\rbr{ R = {1 \over 2B}.}
The action is
\eqn\igb{I = -{\pi Qe^{-2\phi_0} \over 4B} + {\cal{O}}(B^0),}
and the production rate is finite and proportional to
\eqn\eqb{ exp[-{{\pi Q  \over 4B}e^{-2\phi_0}}].}

To go beyond this leading result, one should compute the one-loop
determinant  for whatever quantum fields are present in the theory.
This will include a factor that counts the number of states in the black
hole throat.  However, since the throat is cut off at a finite distance
from the mouth, only a finite number of states will be counted.  Loosely
speaking, there is no pair production of states whose excitations are
localized far down the throat.

Note that our computation actually describes the production of a Wheeler
wormhole rather than an actual pair of monopoles of opposite charge
located inside the throats of growing cornucopions (the picture proposed in
\corn ).  We believe that this is the correct picture of cornucopion
production even in a theory which contains real monopoles, if the
monopoles are sufficiently heavy so that the throat length of the static
extremal black hole is much longer than the distance to the
horizon.
Experiments done over the time scale of
the motion of the cornucopion mouth cannot probe the depths of the
throat
and discover whether or not there are actually monopoles there.  They
cannot distinguish between a Wheeler wormhole and a monopole pair. Our
instanton was contructed by analytic continuation of these Minkowski
trajectories and must therefore conserve magnetic flux without the
intervention of monopoles.  Of course, ordinary monopole pair creation
inside the wormhole can still occur, and it can evolve into the sort of
geometry suggested in \corn .

Let us remind the reader of the limitations on our
semiclassical analysis.  Our considerations are
valid only so long as the horizon forms in a region in which
the coupling is still weak. The coupling at the horizon is given by
$e^{2\phi_H} = {e^{2\phi_0}/ {\gamma B}}$. Thus, if we make the external
field extremely weak, we must also decrease the asymptotic value of
the dilaton in order to stay in the weak coupling regime everywhere in
spacetime.

In summary, the rate of cornucopion pair production in a weak
magnetic field has been estimated in a weakly coupled semiclassical
expansion and found to be finite, despite naive expectations to the
contrary.  Thus an infinite quantum degeneracy of extremal black hole
states may well provide a resolution to the information puzzle for
particle-hole scattering.

While we believe ``infinite volume'' remnants of the general type discussed
here may store the information lost in particle-hole scattering without
implying excessive pair production, it should be stressed that there is little
practical difference between this type of information storage and actual
information loss.  Cornucopions resolve the information puzzle by
forming a new asymptotic region of spacetime.  The $S$-matrix of the
observer in the original asymptotic
region of space (i.e. the region which existed before
cornucopion formation) is not in itself unitary.  If the charged black hole
settles down
into a nonsingular, horizon-free static spacetime, as we have
 been assuming in this paper, then the asymptotic observer could in
 principle construct (two-dimensional!) detectors and send them down the
 cornucopion throat to help him verify the unitarity of the total
 $S$-matrix. Even this will not suffice to extract all information from the
 black hole if the information is receding down the throat at the speed
 of light.  This will
 certainly be the case if the stationary end point of black hole
 evaporation resembles the asymptotically DeSitter solutions of the $N<24$
 semiclassical equations discussed in \Stro .

Thus, in general, our resolution of the information paradox
is in a sense a realization of Hawking's proposal that information
is lost in black hole evaporation.  Note however that this occurs in
the context of a completely unitary quantum mechanical evolution (at least at
the level of the semiclassical approximation for the metric).
Spacetimes like that discussed in \Stro\ can be foliated with a complete
set of Cauchy surfaces, and quantum field theory in such backgrounds is
unitary.  The new feature that allows for information loss is the
creation of a new future asymptotic region of space, causally disconnected
from the region that existed before the formation of the black hole.
The local evolution operator is unitary but the $S$-matrix of the observer
in the initial asymptotic region is not.

It is clear from this discussion that, contrary to Hawking's original
suggestion, virtual processes do not lead to information loss.  Virtual
cornucopion formation and evaporation corresponds to a history in the
path integral in which a finite volume spatial slice of cornucopion is
temporarily formed and then relaxes back to the vacuum.  No new
asymptotic region is formed and no loss of information occurs.  An
effective lagrangian taking into account the effect of small virtual
fluctuations in the geometry will obey the rules of quantum
mechanics, and the difficulties of Hawking's proposal pointed out in
\ref\bps{T. Banks, M. Peskin, L. Susskind,''Difficulties for the evolution
of pure states into mixed states'',\ajou  Nucl. Phys. &B244 (84) 125.} are
avoided.
Our picture of black hole evaporation thus contains Remnants (of
arbitrary states of large black holes) Without Remnants (being produced
copiously in the laboratory or in virtual loops) and Information Loss
(for asymptotic observers) Without Information Loss (in virtual
processes).

Could the scenario proposed here be a general
resolution of the information loss paradox, valid for all processes
involving black holes?  We are not sure.  In the present context,
conservation of magnetic flux prevents the throat of the black hole from
pinching off, creating a disjoint universe into which information can be
truly lost.  It is not obvious what would prevent such
 a process in the case of a neutral black hole.
\vfill\eject
\centerline{\bf Acknowledgements}

We are grateful to  A. Dabholkar,
S. Giddings, G. Horowitz, J. Preskill, N. Seiberg
and S. Shenker for useful conversations.  A. S. would like to thank the
Rutgers Theory Group for their hospitality.  This work was supported in
part by DOE Grants, 91ER40618 and DE-FG-60590-ER-40559.

\listrefs
\end